\documentclass[runningheads]{llncs}
\usepackage{graphicx}
\begin{document}
\title{A Hypergraph-Based Approach to Recommend Online Resources in a Library}
%
%
\author{Debashish Roy\inst{1}\orcidID{0000-1111-2222-3333} \and
Rajarshi Roy Chowdhury\inst{2}\orcidID{0000-0002-1174-3868}}
\authorrunning{D. Roy and RR. Chowdhury}
%
\institute{Toronto Metropolitan University, School of Information Technology Management, Toronto, Canada \\
\email{debashish.roy@torontomu.ca} \and
Metropolitan University, Department of Computer Science and Engineering, Bateshwar, Sylhet, Bangladesh \\
\email{rajarshiry@gmail.com}}
\maketitle              
\begin{abstract}
When users in a digital library read or browse online resources, it generates an immense amount of data. If the underlying system can recommend items, such as books and journals, to the users, it will help them to find the related items. In this research analyzes a digital library's usage data to recommend items to its users, and it uses different clustering algorithms to design the recommender system. We have used content-based clustering, including hierarchical, expectation maximization (EM), K-mean, FarthestFirst and density-based clustering algorithms, and user access pattern-based clustering, which uses a hypergraph-based approach to generate the clusters. This research shows that the recommender system designed using the hypergraph algorithm generates the most accurate recommendation model compared to those designed using the content-based clustering approaches.


\end{abstract}
\section{Introduction}
In this big data era, organizations are investing millions of dollars in understanding the hidden pattern in the dataset. Identifying the hidden pattern helps the organization make profitable decisions for them. According to \cite{Initiative}, Web Analytics uses analytic tools to mine the web usage data retrieved from web servers. Most websites store users' browsing and interaction information in the web server logs to apply web analytics. First, the analytic tools are used to understand user browsing patterns better. Later these patterns are analyzed to increase the usability of the websites. 

An immense amount of data is collected by digital libraries when the users of the library interact with the different online resources the library offers. Therefore, the libraries apply different data analytics tools to analyze the library usage data. By analyzing the usage data, a library can determine the document sets that are least browsed; they can identify the users who use the library the most or the users who do not use the library resources, etc. In this research work, we are analyzing a web server's log. The server logs what items are browsed and saves the resource item's title. Most of the time, we do not have access to the full description of the online resource because of subscriptions and copyright issues. Hence, we have limited text for each server log entry. So, in the recommender system, we have only included the resource item title. The recommender system uses the online resource items of a university library to recommend interesting items to the users of the library. Various content-based and user access pattern-based clustering algorithms are used to design this recommender system. Association rule mining (ARM) is used to generate the user access pattern. We have used a hypergraph-based partitioning algorithm hMETIS to generate clusters from the association rules. We compared each user's browsing history with each generated cluster and calculated Precision, Recall, and F1-Score for each cluster. The result shows that the recommender system generated using the hypergraph partitioning provides the most accurate recommendation.

Furthermore, a careful investigation of the recommended results from the hypergraph-based recommender shows one or more recommended items are content-wise, not similar. But, they are recommended because the recommender uses user access patterns to complete the recommendation. The recommender systems designed using the content-based clustering methods only recommend content-wise similar items. The hypergraph-based recommender system recommends items frequently browsed together to a different user, which is not possible in a content-based recommender system. The targets of this research work are:

\begin{enumerate}
   \item To identify an effective clustering-based recommendation approach for online resources of a library. Here, we compare different clustering algorithms to run the comparison process.
   \item To determine a recommendation approach that will recommend items that are content-wise not similar. Here, the items are content-wise not similar, but they are related because of user access patterns.
   \vspace{-2mm}
\end{enumerate}

The rest of the research paper is structured as follows: \textbf{Section 1.2} focuses on the related works to analyze library dataset, \textbf{Section 1.3} describes the research methodology, \textbf{Section 1.4} discusses the experiment results, and at the end, \textbf{Section 1.5} provides a result summary, and a future direction of this research work.

\section{Related Work}
In recent years, library analytics has been used to help librarians and the users of libraries. Many researchers have proposed different recommender approaches using different characteristics of these datasets, including user access patterns, URL data, and web usage data, for improving the quality of online-based services.    

The research work in \cite{Collins} discusses the use of library usage data to find library usage patterns by undergraduate students. The authors in \cite{Wong} have used library analytics to correlate library material usage with students' academic performance. In reference \cite{Roy-2018}, the authors use association rule-based clustering to cluster the e-resource items of a digital library. Using library usage data, researchers can identify students not studying at the university. In reference \cite{Bridges}, the author is able to group the students who do not study by comparing different undergraduate academic disciplines and library use. Library analytics is used to predict the requirements of future library users. The researchers in reference \cite{Wilson} describe an approach to predict the future usage of a library. In reference \cite{Arendt}, analytics tools are used to provide attractive site development ideas. Standard web analytics tools are not enough to analyze library usage data because the uniform resource locator (URL) pattern for library data is unique. 

Different research works have used different clustering algorithms to group con-tent-wise similar items. Unsupervised classification of the documents is done using clustering algorithms, and different groups are created for similar documents. The authors in reference \cite{Singh} use the K-means clustering algorithm on a term frequency-inverse document frequency (TF-IDF) matrix that represents the resource contents to classify the documents. In their research, they have shown that TF-IDF based clustering provides better results. In reference \cite{Beil}, a text-based clustering algorithm is proposed. In their approach, frequent items are used as terms. Here, frequent itemsets are generated, and later the itemsets are used for association rule mining. The mutual overlaps of the frequent itemsets are used for clustering, and the itemsets are divided into k-partitions. 

In reference \cite{Poongothai}, the authors have compared expectation maximization (EM) clustering, and CFuzzy means for web usage data. Their results show that EM performs 5 to 8 percent better. The authors in reference \cite{Langhnoja} have used the density-based spatial clustering of applications with noise (DBSCAN) clustering algorithm to classify visitor groups with common behavior. They have mined web usage data to ﬁnd similar users. In reference \cite{Han}, the authors have used association rule mining (ARM) based clustering. From the frequent itemsets, they generated diﬀerent association rules. They considered each single association rule as an edge of a hypergraph. So, with all the association rules, a hypergraph is generated. Then for clustering, they apply a hypergraph partitioning algorithm known as hMETIS. 

The authors in reference \cite{Langhnoja-dm} have applied ARM-based clustering for pattern discovery. Here, association rules are formed depending on the frequently browsed pages. To ﬁnd the clusters, they apply the DBSCAN algorithm. ARM-based clustering is also applied to web images to ﬁnd similar types of images \cite{Malik}. The authors generated a hypergraph using all the association rules, and then they applied a hypergraph partitioning algorithm, hMETIS, to generate the clusters. ARM-based clustering is also used to cluster text documents \cite{Liu}. The authors in \cite{Liu} have used the K-means clustering algorithm to cluster the association rules. The authors in reference \cite{Gerardo} have applied hierarchical clustering to design a decision support system. In reference \cite{Jager}, library analytics is applied to ﬁgure out the eﬀect of library usage for successful students. It has been found that students who do poorly in the exam do not use the library resources that much compared to those who do well in the exams. 

Many ongoing research projects are based on library analytics, but none of the projects has become a standard. Researchers from the University of Pennsylvania are designing a data integration framework, Metridoc \cite{Metridoc}, to assist libraries with data ingestion and normalization. The ongoing research works provide usage statistics of different online resource items but does not apply any technique to recommend items based on the clustering technique. This research applies clustering algorithms to recommend online resources to a library user and compares content-based clustering algorithms with usage-based clustering to design a recommender system. For content-based clustering algorithms, we have used density-based clustering, hierarchical clustering, EM clustering, etc. For usage-based clustering, we have used a hypergraph-based association rule clustering algorithm.

\vspace{2mm}
\begin{figure}[htb]
    \centering
    \includegraphics[width = \linewidth]{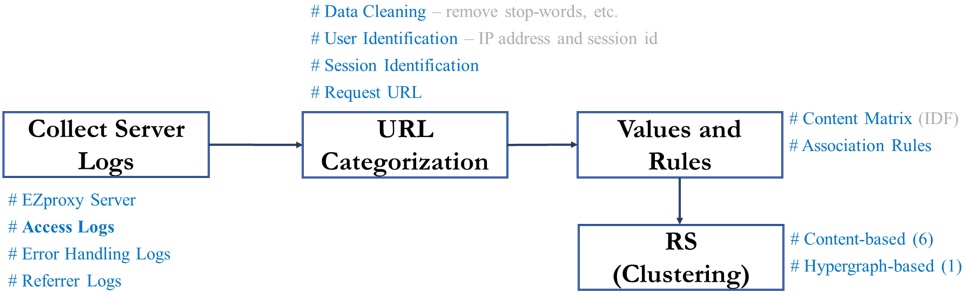}
    \caption{An abstract design of the experimental process.} 
    \label{exp-design}
\end{figure}

\section{Experiment Design}
\begin{figure}[htb]
    \centering
    \includegraphics[width = \linewidth]{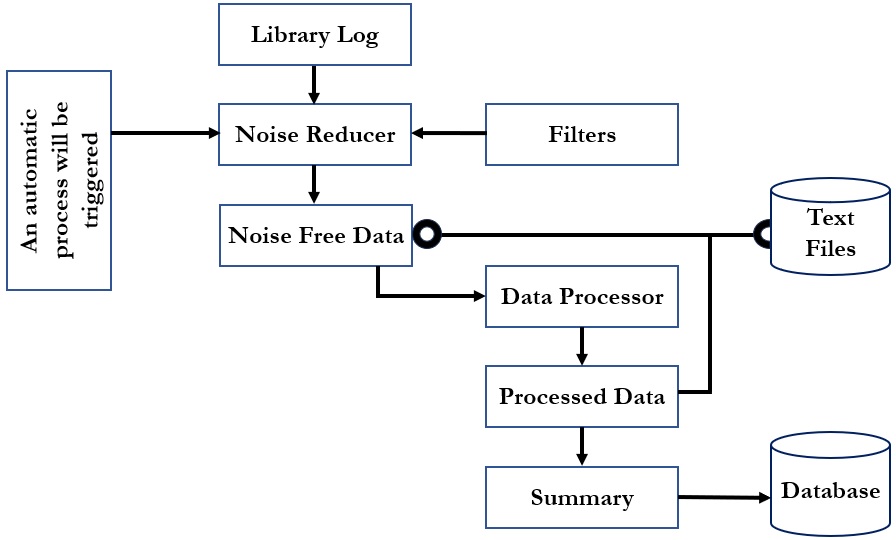}
    \caption{A library Log Analyzer.} 
    \label{A-library-Log-Analyzer}
\end{figure}

This section describes the experiment design step-by-step, as presented in Fig. \ref{exp-design}. At first, server logs are collected from the EZproxy server (it allows libraries to deliver e-resources securely over the network to its authorized users), whilst in this experiment we only collect access logs for analysis. Subsequently, the basic step of machine learning approach, pre-processing, is applied to the collected library server logs to remove the stop-words and unnecessary terms. After data cleaning, a URL categorization is applied based on the users' data and then context matrix and association rules are generated for further processing. Finally, the clustering algorithms are applied to design the recommender system. Fig. \ref{A-library-Log-Analyzer} shows the overall design of the library log analyzer.

\vspace{2mm}

\subsection{Dataset Preparation}
When a user browses the library website, the browsing information is saved in the server log. Several server logs are available such as access-based logs, error-handling logs, agent-handling logs, and referrer logs. We have worked with the access log ﬁles of the library. The library uses the EZproxy server to collect the access logs. EZproxy is a middle-ware solution that the libraries use to provide access to its users who access the library resources outside the regular library network. It stores usage information using a standard web server log ﬁle \cite{OCLC}. In our case, EZproxy collects usage information for the university using a common log format. EZproxy records usage information using a common log format. EZproxy server for the library uses the following log format:
\newline
\begin{center} \%h \%u \%l \%{ezproxy-session} i \%t "\%r" \%s \%b  \end{center}
\hfill \break
Here, \%h is the host accessing the EZproxy server and it is always an IP address, \%u is the username used to log into the EZproxy server, \%l is the remote username but it is always a -, \%{ezproxy-session} i is EZproxy's identifier for the user's current session, \%t captures the timestamp, \%r gets the complete request with server information, \%s collects the status codes returned by the server and \%b provides a statistic number of bytes transferred. Here is an example log entry created by the EZproxy server: \\ \newline

$\backslash$10.0.0.1 X2bFdM1R3txwlkv - 13d8f72f08d1a4e1c418a7cb8fc31437[01/Jun/2014: 00:47:10-0500] $\backslash$GET http:// site.ebrary.com:80/lib/oculryerson/docDetail.\\ action?docID=10251051 HTTP/1.1" 200 29732" 
\newline
\\
The entry says user `X2bFdM1R3txwlkv' from IP address '10.0.0.1' having a session-id`13d8f72f08d1a4e1c418a7cb8fc31437' requested to read a document having id '10251051' on the library server at the given time stamp and the status code 200 says, the requested resource was found on the server so 29732 bytes were transferred to the user.

EZproxy creates a log ﬁle entry each time an e-resource item is accessed through the EZproxy server. EZproxy logs each ﬁle downloaded by a user. It also logs the search queries made against the databases. When we browse a web page, a hypertext markup language (HTML) ﬁle and diﬀerent images, scripts, etc., are downloaded. Requests are sent to the server to download each item. As the server works to serve the request, it also creates an entry in the log ﬁle. Although the user downloads only one ﬁle, multiple lines of entries are written to the server log. In the cleaning process, we must remove all additional requests the users do not request. To clean the log ﬁles, we removed all the URLs with suﬃxes like jpeg, jpg, gif, CSS, etc. We have also worked with the status codes returned by the servers to do the cleaning. We can determine whether the submitted request is successful by examining the status code. As the status code of successful requests ranges from 200-299 \cite{Fielding},  we have removed all the entries with status codes outside 200-299.

\subsection{User and Session Identification with URL Extraction}
After cleaning the log ﬁle, we start the process of user identiﬁcation. Using the EZproxy log, we can identify a user by the internet protocol (IP) address and user id. In our architecture, we have identiﬁed the user using both IP address and user id. After user identiﬁcation, the next step is to identify the sessions created by diﬀerent users. User session identiﬁcation from an EZproxy server log is relatively easy because the log ﬁle stores the session id. So, using the time stamp associated with the session id, we can calculate the session length. A university library has subscriptions to several research databases. The databases have their own URL patterns to access resources such as journals, articles, e-books, etc. Following is an example of a resource request 
\\ \newline
\textbf{URL}:\textit{ site.ebrary.com:80/lib/oculryerson/docDetail.action?do-cID=10017060}. 
\\ \newline
Diﬀerent vendors have different naming patterns for these URLs, but using our system, we can retrieve article information such as journal id, page numbers, volume numbers, etc., from diﬀerent resource request URLs.

\subsection{Recommender System using Online Resource Items}
This experiment uses words from each document title to form a dictionary. Then, a matrix of inverse document frequency (IDF) values is constructed for each word of the individual title. For IDF calculation, we have used the following:


\begin{equation}
Idf_k = log(\frac{N}{df_k})
\end{equation}

Here, $df_k$ denotes the number of documents that consist of the $k^th$ term, and $N$ is used for the total number of documents in the dictionary.

Now that the content matrix is ready, content-based clustering algorithms are applied to design the recommender system. Next, to apply hypergraph-based clustering, we generate the association rules. We use association rules to identify which online items are accessed together by a user in a session. We use this information to understand better how the library resources are being used and how the resources are correlated through users' eyes. Association rule is expressed by the following:


\begin{equation}
(X1, X2) = \Rightarrow X3)
\end{equation}

This rule means \textit{X1} and \textit{X2} imply \textit{X3}, i.e., if \textit{X1}, \textit{X2}, and \textit{X3} are three articles, then any user who reads articles \textit{X1} and \textit{X2} has a good chance that the user will read article \textit{X3}. To measure the chance of the appearance of article \textit{X3}, association rules provide two important metrics: support and conﬁdence. Support and conﬁdence are two user deﬁned parameters. The user sets minimum support and minimum conﬁdence to generate the most frequent itemsets. For example, if the support of an association rule is 10\%, it means 10\% of the total records in the database contain this rule. The association rules that have lower support are not frequent in the database, so rules having lower support could be ignored by the users. Another important parameter for association rule mining is conﬁdence. For example, if the conﬁdence for the association rule ${(X1, X2) \Rightarrow X3}$ is 90\%, then it means 90 percent of the records that contain (\textit{X1}, \textit{X2}) also contain \textit{X3}. In our research, we have set support to a very low value of 1\% because we want to ﬁnd out surprise relationships among diﬀerent documents. And to get the strong rules, we have set conﬁdence to 80\%. All the association rules are used to generate a hypergraph. In a hypergraph, there are vertices and hyperedges connecting the vertices. In this experiment, we have used each association rule to create a hyperedge. The hyperedge vertices are mapped to the documents of the library. hMETIS is a hypergraph partitioning tool that can generate balanced \textit{k}-partitions. So, for hypergraph partitioning, we have used the standalone tool hMETIS.

\subsection{Tools Used}
This experiment uses a machine-learning tool WEKA \cite{Frank} to generate the association rules. WEKA is a tool that houses a set of machine-learning algorithms to solve different data mining problems, including classification \cite{Chowdhury-2022}, \cite{Chowdhury-2023}, \cite{Chowdhury-IAES}, \cite{Chowdhury-sh-iot}, \cite{Ortega-ijatcse},\cite{Chowdhury-turkish}, \cite{Aneja}, \cite{Chowdhury-uob}, clustering \cite{Roy-2023}, \cite{Dol-clus}, \cite{Saputra-clus}, and association \cite{Mohamed}, cite{Naajim} tasks. This state-of-the-art software can be used either using command line interface (CLI) or graphical user interface (GUI) interfaces. hMETIS is a standalone application used to partition the hypergraph. The experiment is implemented using the programming language Java.

\section{Results Analysis}
This experiment applies different clustering algorithms to design a recommender system. For result analysis, we have compared the recommendation accuracy for the recommender systems generated using the clustering algorithms. To start the clustering process, first, we selected the top 254 browsed documents. Here, we selected the documents that were browsed at least ten times. Next, we apply the clustering algorithms to generate seventeen clusters. Here, we have chosen seventeen because, with the librarian's help, we manually grouped the titles into a total of seventeen (17) groups.

To continue with the design of the recommender system, we extracted 353 user profiles, where a profile consists of the browsed documents by the users. Here, to select the 353 users, we have selected the users who have at least browsed fifteen (15) items. Next, we compare the user profiles with each of the seventeen clusters generated by the algorithms. For each comparison, we calculate the Precision (\textit{P}), Recall (\textit{R}), and F1-Score. These metrics quantify the effectiveness of the proposed recommender system using a different set of machine learning algorithms (ML). We have used the following to calculate Precision, Recall, and F1-Score:

\begin{equation}
P = \frac{Number ~ of ~ correctly ~ classified ~ items ~ that ~ belong ~ to ~ the ~ positive ~ class}{Total ~ number ~ of ~ items ~ classified ~ that ~ belong ~ to ~ the ~ positive ~ class}
\end{equation}

\begin{equation}
R = \frac{Number ~ of ~ correctly ~ classified ~ items ~ that ~ belong ~ to ~ the ~ positive ~ class}{Total ~ number ~ of ~ elements ~ that ~ actually ~ belong ~ to ~ the ~ positive ~ class}
\end{equation}

\begin{equation}
F1-Score = \frac{2 \times Precision \times Recall}{Precision + Recall}
\end{equation}

Each user profile is compared with seventeen clusters, so one user profile generates seventeen sets of {Precision, Recall, F1-Score}. We calculate the highest of the seven-teen scores to get a single set of {Precision, Recall, F1-Score} for each user. In this way, for each clustering algorithm, we get 353 sets of {precision, recall, F1-Score}. Finally, we average them to get a single set of {Precision, Recall, F1-Score} for each clustering algorithm. Table \ref{performances} shows the results of the experiment for different clustering algorithms.

\begin{table}
\begin{center}
\caption{Precision, Recall and F1-Score for the clustering algorithms.}\label{performances}
\begin{tabular}{|l|l|l|l|}
\hline
Algorithm &  Precision & Recall & F1-Score \\ 
\hline
EM & 0.6 & 0.34 & 0.43 \\
Filtered & 0.56 & 0.4 & 0.47  \\ 
K-Mean & 0.65 & 0.45 & 0.53  \\ 
FarthestFirst & 0.4 & 0.33 & 0.36  \\ 
Density & 0.52 & 0.38 & 0.44  \\ 
Hierarchical & 0.45 & 0.32 & 0.37  \\ 
Hypergraph & 0.75 & 0.6 & 0.67  \\
\hline
\end{tabular}
\end{center}
\end{table}

Table \ref{performances} illustrates that the hypergraph-based recommender system generates the most accurate recommendations compared to other algorithms. For example, the hypergraph-based recommender recommends items with 75\% Precision, and also it has the highest F1-Score. Furthermore, out of the content-based clustering algorithms, the recommender performs best with the K-mean clustering algorithm with a precision of 65\% and an F1-Score of 0.53.

This section has analyzed the experiment results. We have used different clustering algorithms to design the recommender system. We have found that the hypergraph-based recommender system generates the most accurate recommendations from all the clustering algorithms.

\section{Conclusion and Future Work}
In this research work, we have worked with library usage data retrieved from the web server log. We have designed a recommender system that uses library usage data. The recommender system is designed based on the documents' titles and users' resource access patterns. The experiment result shows that the hypergraph-based recommender system outperforms (precision 75\%) all other recommender systems generated using the content-based clustering algorithms, whilst the FarthestFirst algorithm gains the lowest performances compared to all other algorithms. In terms of the content-based clustering approach, the K-mean algorithm shows high accuracy.    

In the future, we aim to compare this hypergraph-based recommender system with matrix factorization and deep neural network-based recommender systems. So, we will change our experiment design to include matrix-factorization and deep neural networks. \newline
\\
\textbf{Acknowledgement} \\
The authors are profoundly grateful to the Toronto Metropolitan University, Ontario, Canada, and the Metropolitan University (MU), Pirer bazar, Bateshwar, Sylhet-3104, Bangladesh, for supporting this research work. \newline
\\
\textbf{Conflict of Interest}\\
There are no conflicts of interest from the authors.



\begin{thebibliography}{8}

\bibitem{Initiative}
Initiative, L. E. 7 Things you should know about ANALYTICS.URL: \url{https://library.educause.edu/-/media/files/library/2010/4/eli7059-pdf.pdf} (2010).

\bibitem{Collins}
Collins, E. \& Stone, G. Understanding Patterns of Library Use Among Under-graduate Students from Different Disciplines. {\em Evidence-Based Library and Information Practice}. \textbf{9(3)} pp. 51-67 (2014).

\bibitem{Wong}
Wong, S. \& Webb, T. Uncovering meaningful correlation between student academic performance and library material usage. {\em Association of College \& Research Libraries}. (2010).

\bibitem{Roy-2018}
Roy, D., Ding, C., Jin, L. \& Thomas, D. Association rule-based clustering of electronic resources in University Digital Library. {\em In Digital Libraries for Open Knowledge: 22nd International Conference on Theory and Practice of Digital Libraries}. (2018).

\bibitem{Bridges}
Bridges, L. Who is not using the library? A comparison of undergraduate academic disciplines and library use. {\em The Johns Hopkins University Press, Portal: Libraries and the Academy}. \textbf{8(2)} pp. 187-196 (2008).

\bibitem{Wilson}
Wilson, M. Understanding the needs of tomorrow's library user: Rethinking library services for the new age. {\em Australasian Public Libraries and Information Services}. \textbf{13(2)} p. 81 (2000).

\bibitem{Arendt}
Arendt, J. \& Wagner, C. Beyond Description: Converting Web Site Usage Statistics into Concrete Site Improvement Ideas. {\em Journal of Web Librarianship}. \textbf{4(1)} p. 3754 (2000).


\bibitem{Singh}
Singh, V.K., Tiwari, N. \& Garg, S. Document Clustering using K-Means, Heuristic K-Means and Fuzzy C-Means. {\em International Conference on Computational Intelligence and Communication Networks (CICN)}. pp. 297-301 (2011).

\bibitem{Beil}
Beil, F., Ester, M. \& Xu, X. Frequent term-based text clustering. {\em In the Proceedings of the eighth ACM SIGKDD international conference on Knowledge discovery and data mining}. pp. 436-442 (2002).

\bibitem{Poongothai}
Poongothai, K., Parimala, M. \& Sathiyabama, S. Efficient Web Usage Mining with Clustering. {\em International Journal of Computer Science}. \textbf{8(6)} pp. 203-209 (2011).

\bibitem{Langhnoja}
Langhnoja, S. G., Mehul, P. B. \& Darshak, B. M. Web Usage Mining to Discover Visitor Group with Common Behavior Using DBSCAN Clustering Algorithm. {\em International Journal of Engineering and Innovative Technology (IJEIT)}. \textbf{2} pp. 169-173 (2013).

\bibitem{Han}
Han, E., Karypis, G. \& Mobasher, B. Clustering Based On Association Rule Hypergraphs. {\em Workshop on Research Issues on Data Mining and Knowledge Discovery}. pp. 9-13 (1997).

\bibitem{Langhnoja-dm}
Langhnoja, S. G., Mehul, P. B. \& Darshak, B. M. Web Usage Mining Using Association Rule Mining on Clustered Data for Pattern Discovery. {\em International Journal of Data Mining Techniques and Applications}. \textbf{2(1)} pp. 141-150 (2013).

\bibitem{Malik}
Malik, H. \& Kender R. Clustering Web Images using Association Rules, Interestingness Measures, and Hypergraph Partitions. {\em Proceedings of the 6th International Conference on Web Engineering, ACM}. pp. 48-55 (2006).

\bibitem{Liu}
Liu, G., Huang, S., Lu, C. \& Du, Y. An improved K-Means Algorithm Based on Association Rules. {\em International Journal of Computer Theory and Engineering (IJCTE)}. \textbf{6(2)}, pp. 146-149 (2014).

\bibitem{Gerardo}
Gerardo, B. D., Byun, Y. \& Tanguilig III, B. Hierarchical Clustering and Association Rule Discovery Process for Efficient Decision Support System. {\em Communication and Networking, Communications in Computer and Information Science}. \textbf{266} pp. 239-247 (2011).

\bibitem{Jager}
Jager, K. d. Successful students: does the library make a difference? {\em Performance Measurement and Metrics}. \textbf{3(3)} pp. 140-144 (2002).

\bibitem{Metridoc}
Metridoc. Penn Library DataFarm: Metridoc Project. {\em University of Pennsylvania}. URL: \url{https://metridoc.library.upenn.edu/} (2023).

\bibitem{OCLC}
OCLC. EZproxy Conﬁguration: LogFormat. URL: \url{https://www.oclc.org/support/services/ezproxy/documentation/cfg/logformat.en.html}\\(2023).

\bibitem{Fielding}
Fielding, R. \& Adobe, E. Hypertext Transfer Protocol (HTTP/1.1): Semantics and Content. {\em Internet Engineering Task Force}. URL: \url{https://www.rfc-editor.org/rfc/rfc7231\#page-47} (2023).

\bibitem{Frank}
Frank, E., Hall, M. A. \& Witten, I. H. The WEKA Workbench. Online Appendix for Data Mining: Practical Machine Learning Tools and Techniques, 4th ed. Morgan Kaufmann. URL: \url{https://www.cs.waikato.ac.nz/ml/weka/Witten\_et\_al\_2016\_appendix.pdf} (2023).

\bibitem{Chowdhury-2022}
Chowdhury, R. R. \& Abas, P. E. A survey on device fingerprinting approach for resource-constraint IoT devices: Comparative study and research challenges. {\em Internet of Things (Netherlands)}. \textbf{20} (2022). 

\bibitem{Chowdhury-2023}
Chowdhury, R. R., Idris, A. C. \& Abas, P. E. A Deep Learning Approach for Classifying Network Connected IoT Devices Using Communication Traffic Characteristics. {\em Journal of Network and Systems Management}. \textbf{31(1)} p. 26 (2023).

\bibitem{Chowdhury-IAES}
Chowdhury, R. R., Idris, A. C. \& Abas, P. E. Device identification using optimized digital footprints. {\em IAES International Journal of Artificial Intelligence}. \textbf{12(1)} pp. 232–240 (2023).

\bibitem{Roy-2023}
Roy, D., Chowdhury, R. R., Nasser, A. B., Azim, A. \& Babaeianjelodar, M. Item recommendation using user feedback data and item profile. {\em AIP Conference Proceedings}. \textbf{2643} (2023).

\bibitem{Mohamed}
Mohamed, K. \& Bayraktar, Ü. A. Artificial Intelligence in Public Relations and Association Rule Mining as a Decision Support Tool. {\em International Journal of Humanities and Social Science}. \textbf{9(3)} pp. 23-32 (2022).

\bibitem{Chowdhury-sh-iot}
Chowdhury, R. R., Idris, A. C. \& Abas, P. E. Identifying SH-IoT devices from network traffic characteristics using random forest classifier. {\em Wireless Networks}. (2023).

\bibitem{Dol-clus}
Dol S. M. \& Jawandhiya P. M. Classification Technique and its Combination with Clustering and Association Rule Mining in Educational Data Mining — A survey. {\em Engineering Applications of Artificial Intelligence}. \textbf{122} p. 106071 (2023). 

\bibitem{Saputra-clus}
Saputra D., Haryani, Junaidi A., Baidawi T. \& Surniandari A. Application of K-mean clustering algorithm in grouping data prospective new students. {\em 2ND INTERNATIONAL CONFERENCE ON ADVANCED INFORMATION SCIENTIFIC DEVELOPMENT (ICAISD) 2021: Innovating Scientific Learning for Deep Communication}. (2021).

\bibitem{Ortega-ijatcse}
Ortega, J. H. J. C., Resureccion, M. R., Natividad, L. R. Q., Bantug, E. T., Lagman, A. C. \& Lopez, S. R. An Analysis of Classification of Breast Cancer Dataset Using J48 Algorithm. {\em International Journal of Advanced Trends in Computer Science and Engineering}. \textbf{9(1.3)} (2020).

\bibitem{Chowdhury-turkish}
Chowdhury, R. R., Idris, A. C. \& Abas, P. E. Packet-level and IEEE 802.11 MAC frame-level analysis for IoT device identification. {\em Turkish Journal of Electrical Engineering and Computer Sciences}. \textbf{30(5)} pp. 1941–1961 (2022).

\bibitem{Aneja}
Aneja, S., Bhargava, B. K., Aneja, N. \& Chowdhury, R. R. Device Fingerprinting using Deep Convolutional Neural Networks. {\em International Journal of Communication Networks and Distributed Systems}. \textbf{28(2)} pp. 171–198 (2022).

\bibitem{Naajim}
Naajim, M., Vickramkarthick, Radhakrishnan, Jatain, A. Association Rules Generation for Injuries in National Football League (NFL). Proceedings of International Conference on Advanced Communications and Machine Intelligence. MICA 2022. Studies in Autonomic, Data-driven and Industrial Computing. Springer, Singapore. (2023).

\bibitem{Chowdhury-uob}
Chowdhury, R. R., Idris, A. C. \& Abas, P. E. Internet of Things Device Classification using Transport and Network Layers Communication Traffic Traces. {\em International Journal of Computing and Digital Systems}. \textbf{12(1)} pp. 2210–142 (2022).
   

\end{thebibliography}
\end{document}